\documentclass[aps,prl,twocolumn,showpacs,preprintnumbers,amsmath,amssymb,superscriptaddress]{revtex4}

\usepackage{graphicx}% Include figure files
\usepackage{dcolumn}% Align table columns on decimal point
\usepackage{bm}% bold math
\newcommand{\nc}{\newcommand}
\nc{\be}{\begin{equation}}
\nc{\ee}{\end{equation}}
\nc{\bea}{\begin{eqnarray}}
\nc{\eea}{\end{eqnarray}}
\nc{\bean}{\begin{eqnarray*}}
\nc{\eean}{\end{eqnarray*}}
\nc{\mb}{\mbox}
\nc{\rnc}{\renewcommand}
\nc{\vk}{\mb{\bf k}}
\nc{\vp}{\mb{\bf p}}
\nc{\vn}{\mb{\bf n}}
\nc{\vq}{\mb{\bf q}}
\nc{\rr}{\mb{\bf r}}
\nc{\vz}{\hat {\mb{\bf z}}}
\nc{\vj}{\mb{\boldmath$j$}}
\nc{\vg}{\mb{\boldmath$g$}}
\nc{\x}{\mb{\boldmath$x$}}
\nc{\A}{\mb{\boldmath$A$}}
\nc{\va}{\mb{\boldmath$a$}}
\nc{\vs}{\mb{\boldmath$\sigma$}}
\nc{\vpi}{\mb{\boldmath$\pi$}}
\nc{\nab}{\nabla}
\nc{\X}{\sf x}

\begin{document}

\title{The Role of Electron-electron Interactions in Graphene ARPES Spectra}

\author{Marco Polini}
\email{m.polini@sns.it}
\affiliation{NEST-CNR-INFM and Scuola Normale Superiore, I-56126 Pisa, Italy}
\author{Reza Asgari}
\affiliation{Institute for Studies in Theoretical Physics and Mathematics, Tehran 19395-5531, Iran}
\author{Giovanni Borghi}
\affiliation{NEST-CNR-INFM and Scuola Normale Superiore, I-56126 Pisa, Italy}
\author{Yafis Barlas}
\affiliation{Department of Physics, The University of Texas at Austin, Austin Texas 78712}
\author{T. Pereg-Barnea}
\affiliation{Department of Physics, The University of Texas at Austin, Austin Texas 78712}
\author{A.H. MacDonald}
\affiliation{Department of Physics, The University of Texas at Austin, Austin Texas 78712}

\begin{abstract}
We report on a theoretical study of the influence of electron-electron interactions 
on ARPES spectra in graphene that is based on the random-phase-approximation and on graphene's
massless Dirac equation continuum model.  We find that level repulsion between quasiparticle and 
plasmaron resonances gives rise to a gap-like feature at small $k$.  
ARPES spectra are sensitive to the electron-electron
interaction coupling strength $\alpha_{\rm gr}$ and might enable an experimental determination of this
material parameter.
\end{abstract}


\maketitle

\noindent
{\em Introduction}---
In quasi two-dimensional (2D) crystals ARPES is a powerful probe of band structure 
and of Coulomb and phonon-mediated electron-electron interactions.
A number of recent experiments~\cite{arpes_expt} have reported ARPES spectra for
single-layer and bilayer graphene systems grown epitaxially~\cite{deHeerSSC} on the surface of  
SiC.  These recently isolated~\cite{geim} 2D crystals have attracted 
considerable attention because of unusual properties~\cite{novoselov,PT}
that follow from chiral band states, notably unusual quantum Hall effects~\cite{QHE},
and because of potential for applications.
Experiment appears to establish~\cite{arpes_expt} that both electron-phonon and electron-electron
dressings have an influence on graphene's ARPES spectra.  Recent theoretical studies~\cite{mauri,louie,dassarma_eph}
provide a solid basis for interpreting the electron-phonon contributions; in this Letter  
we report on a theoretical study of the influence of electron-electron interactions.
Our analysis is based on the massless Dirac model which describes~\cite{andoDirac} the $\pi$ electron states within
$\sim 2 {\rm eV}$ of the Fermi energy of a neutral graphene sheet and on the random phase approximation
(RPA) for the self-energy.

\begin{figure}[t]
\begin{center}
\tabcolsep=0cm
\includegraphics[width=1.00\linewidth]{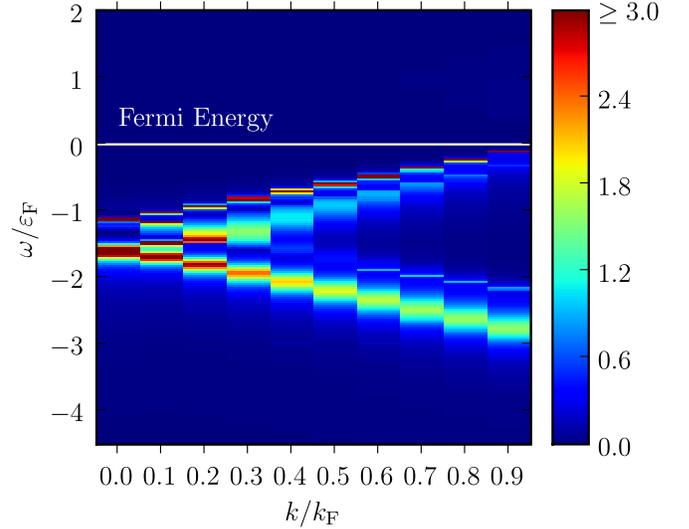}
\caption{(Color online) Spectral function ${\cal A}({\bm k},\omega)$ of a n-doped graphene sheet for 
with wavevector $k$ in units of Fermi wavevector $k_{\rm F}$ 
and energy $\omega$ in units of and measured from the Fermi energy $\hbar v k_{\rm F}$.  The ARPES spectra for each 
$k$ is the portion of the spectral function with $\omega < 0$.  The $k$-dependence is represented in this figure by 
results for ten discrete $k \in [0.0,0.9]$ values separated by $0.1$.  It follows from particle-hole symmetry that the spectral 
function of a p-doped graphene sheet is specified by the $\omega > 0$ spectral weight.
\label{fig:one}}
\end{center}
\end{figure}

%Marco: added ``dynamically screened" because SX term in SX-CH self-energy decomposition has v_q --> v_q/\varepsilon(q,\omega)
The self-energy in a system of fermions is naturally separated into an
exchange contribution due to interactions with occupied states in the static Fermi sea,
and a correlation contribution due to quantum fluctuations of the Fermi sea~\cite{Giuliani_and_Vignale}.  
Graphene differs~\cite{poliniSSC}  
from the widely studied 2D systems in semiconductor quantum wells because its quasiparticles
are chiral and because it is gapless and therefore has interband quantum fluctuations on the Fermi energy 
scale.  In graphene band eigenstate chirality endows exchange interactions with a new source of momentum dependence 
which renormalizes the quasiparticle velocity and strongly
influences the compressibility and the spin-susceptibility~\cite{diracgaspapers,ourprl,dassarmaselfenergy}.

Our results~\cite{results} for the ARPES spectra of a n-doped graphene sheet with $\alpha_{\rm gr}=2$
are summarized in Fig.~\ref{fig:one}.  Here $\alpha_{\rm gr}=g e^2/\epsilon \hbar v$ 
is the interaction coupling strength, $g=g_{\rm spin}g_{\rm valley}=4$ is a band degeneracy factor, 
$v$ is the Fermi velocity, and $\epsilon$ depends on the dielectric environment of the graphene layer 
($\alpha_{\rm gr} \sim 2$ is a typical value thought to apply to graphene sheets on the surface of a ${\rm SiO}_2$ substrate). 
The most notable aspect of these results is the appearance near $k=0$ of strong 
plasmaron peaks in addition to quasiparticle peaks, which give rise to a gap-like 
structure in the overall spectrum.  In the following
paragraphs we explain the physics behind this figure.

\begin{figure}[t]
\begin{center}
\tabcolsep=0cm
\begin{tabular}{cc}
\includegraphics[width=0.50\linewidth]{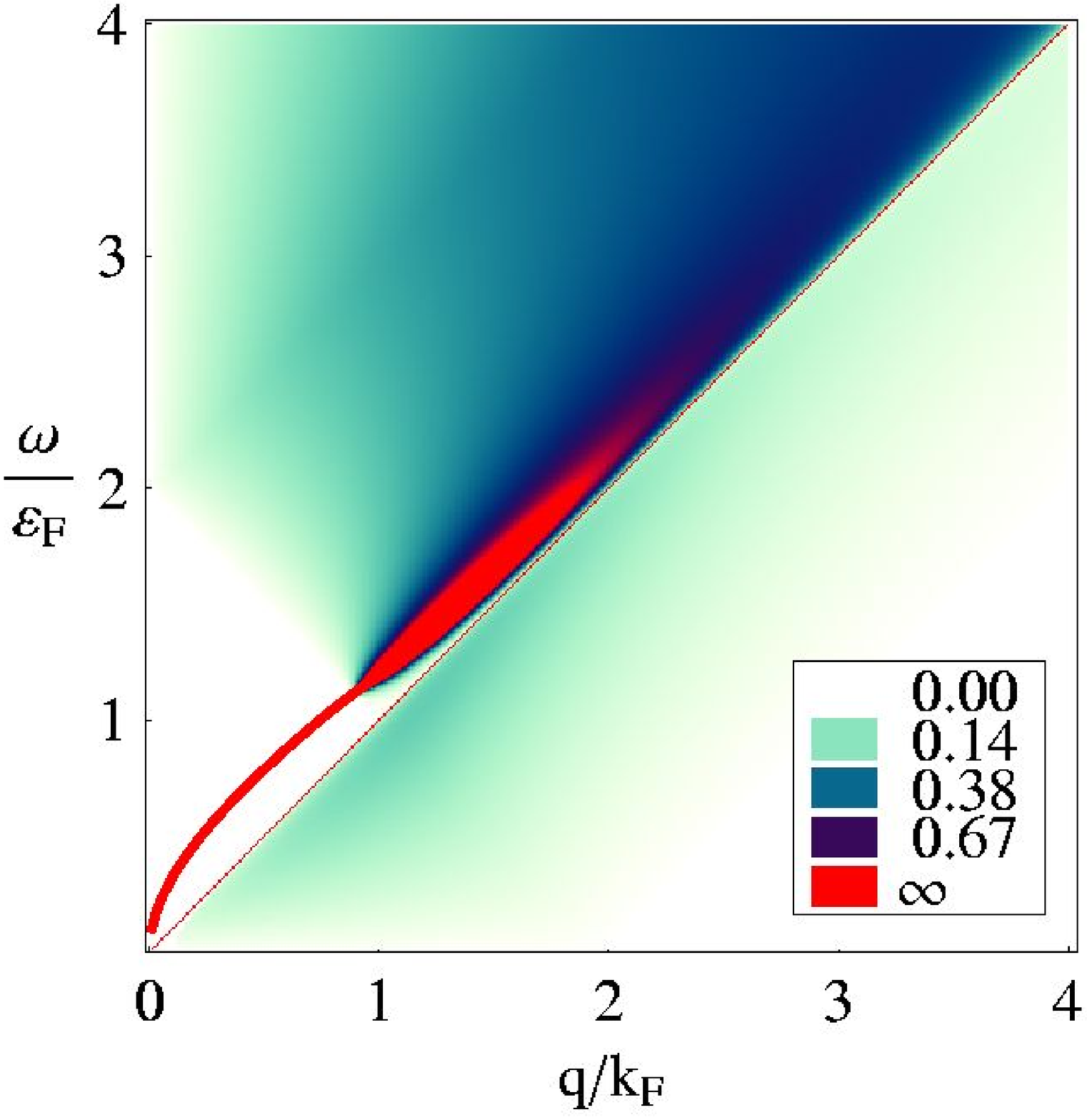}&
\includegraphics[width=0.50\linewidth]{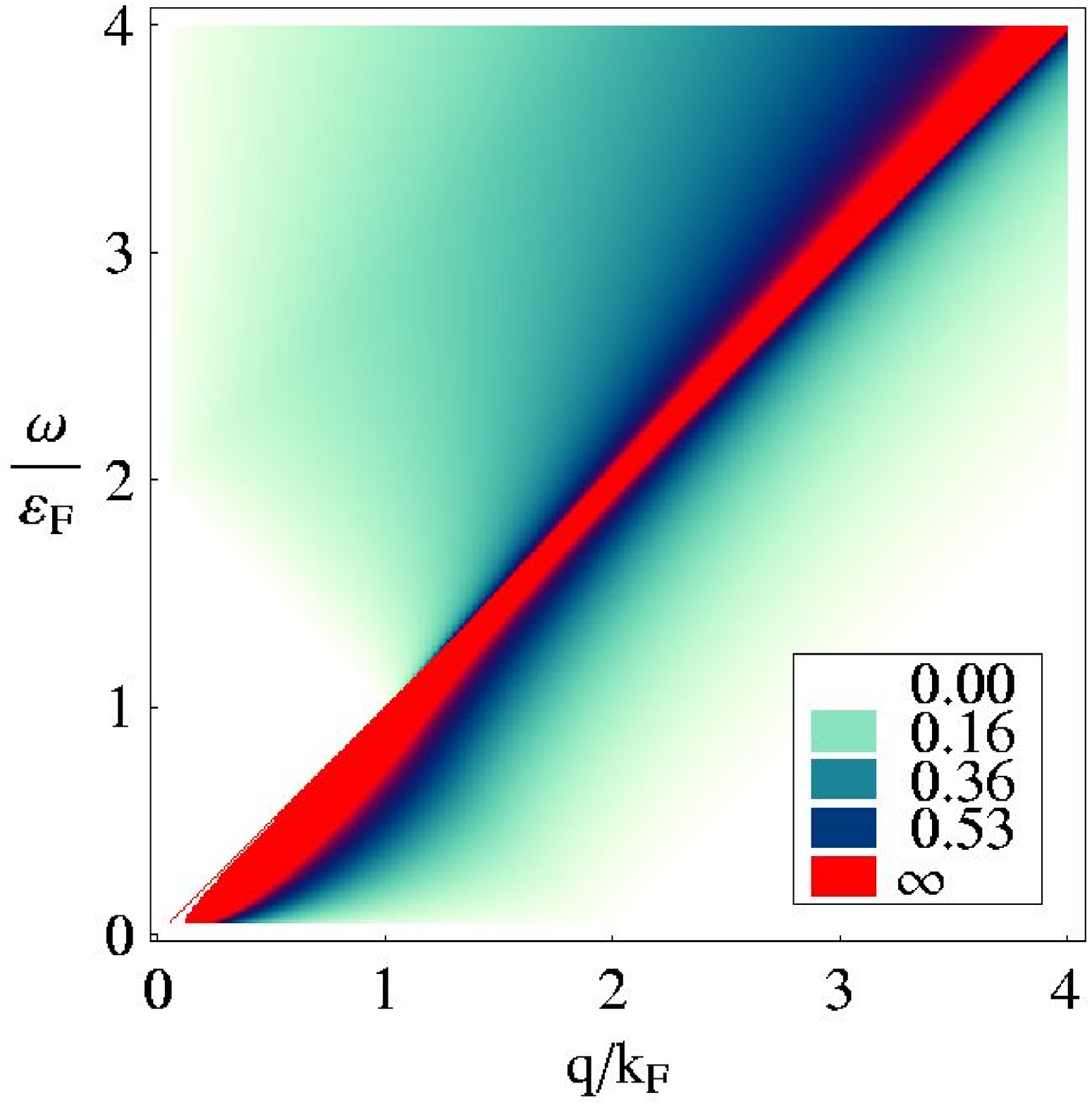}\\
\end{tabular}
\caption{(Color online) Left panel: $-\Im m [\varepsilon^{-1}({\bm q}, \omega)]$ as a function of $q/k_{\rm F}$ and  $\omega/\varepsilon_{\rm F}$ for $\alpha_{\rm gr}=2$. The red solid line is the plasmon dispersion relation, $\omega_{\rm pl}(q \to 0)=\varepsilon_{\rm F}\sqrt{\alpha_{\rm gr} q/(2 k_{\rm F})}$. Right panel: $-v_q \Im m [\chi^{(0)}({\bm q}, \omega)]$ as a function of $q/k_{\rm F}$ and  $\omega/\varepsilon_{\rm F}$.
The left and right panels become identical in the non-interacting $\alpha_{\rm gr} \to 0$ limit.\label{fig:two}}
\end{center}
\end{figure}

\noindent
{\em Doped Dirac Sea Charge Fluctuations}---
The massless Dirac band Hamiltonian of graphene is~\cite{PT}
\begin{equation}\label{eq:Dirac}
{\cal H}= v \tau \left(\sigma_1 \, p_1 + \sigma_2 \, p_2 \right)\,,
\end{equation}
where $\tau = \pm 1$ for the inequivalent $K$ and $K'$ valleys at which $\pi$ and $\pi^*$ bands 
touch, $p_i$ is an envelope function
momentum operator, and $\sigma_i$ is a Pauli matrix which acts on the sublattice pseudospin
degree-of-freedom.  The low-energy valence band states have pseudospin aligned with momentum,
while the high energy conduction band states, split by $2 v |\bm{p}|$, are anti-aligned.  In 
Fig.~\ref{fig:two} we compare the particle-hole excitation~\cite{ourprl,Lindhard} spectra
of non-interacting and interacting 2D doped Dirac systems.
The non-interacting particle-hole continuum is represented here by
the imaginary part of graphene's Lindhard function, $\Im m[\chi^{(0)}({\bm q},\omega)]$, 
which weighs transitions by the strength of the density fluctuation to which they
give rise; transitions between states with opposite pseudospin  
orientation therefore have zero weight.  More generally the band-chirality related density-fluctuation weighting factor
(called the chirality factor below), which plays a key role in the physics of the ARPES spectra, is
$[1 \pm \cos(\theta_{{\bm k}, {\bm k}+{\bm q}})]/2$ with the plus sign applying for intraband 
transitions and the minus sign applying for interband transitions, and $\theta_{{\bm k}, {\bm k}+{\bm q}}$ equal to the angle between 
the initial state (${\bm k}$) and final state (${\bm k}+{\bm q}$) momenta.  
The weight is therefore high for intraband (interband) transitions when ${\bm k}$ and ${\bm k}+{\bm q}$ are in the same 
(opposite) direction.  The main features in Fig.~\ref{fig:two} that are important for ARPES spectra are the $1/\sqrt{vq-\Omega}$ 
divergence which occurs near the upper limit of the $q < k_{\rm F}$ intraband particle-hole continuum and the 
relatively weak weight at the lower limit of the $q < k_{\rm F}$ inter-band particle-hole continuum.  The divergence at the
intraband particle-hole spectrum contrasts with the singular but finite $\sqrt{\Omega_{\rm max}-\Omega}$ 
behavior at the upper end of the particle-hole continuum in an ordinary electron gas.  The difference follows from the linear quasiparticle dispersion
which places the maximum intraband particle-hole excitation energy at $v q$ for all $k$ in the Fermi sea.

In the RPA, quasiparticles interact with Coulomb-coupled particle-hole excitations.  
Because the bare particle-hole excitations are more sharply 
bunched in energy, Coulomb coupling leads to plasmon excitations 
that are sharply defined out to larger wavevectors than in the ordinary electron gas and 
steal more spectral weight from the particle-hole continuum.  As seen in Fig.~\ref{fig:two}, the plasmon
excitation of the Dirac sea remains remarkably well defined even when it enters the interband particle-hole
continuum.  The persistence occurs because transitions near the bottom of the interband particle-hole 
continuum have nearly parallel ${\bm k}$ and ${\bm k}+{\bm q}$ and therefore little charge-fluctuation weight.
Interactions between quasiparticles and plasmons are stronger in the 2D massless Dirac system than in an ordinary non-relativistic 2D system.

\begin{figure}[t]
\begin{center}
\tabcolsep=0cm
\begin{tabular}{cc}
\includegraphics[width=0.50\linewidth]{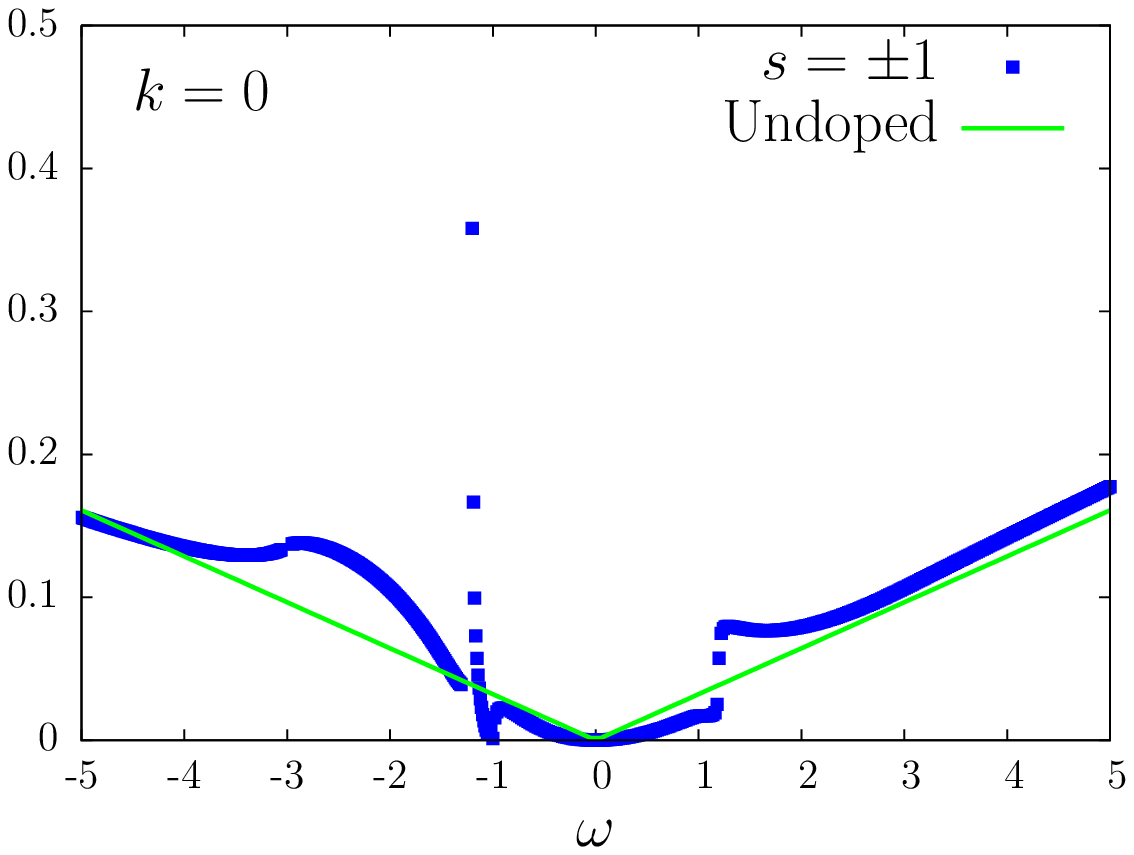}&
\includegraphics[width=0.50\linewidth]{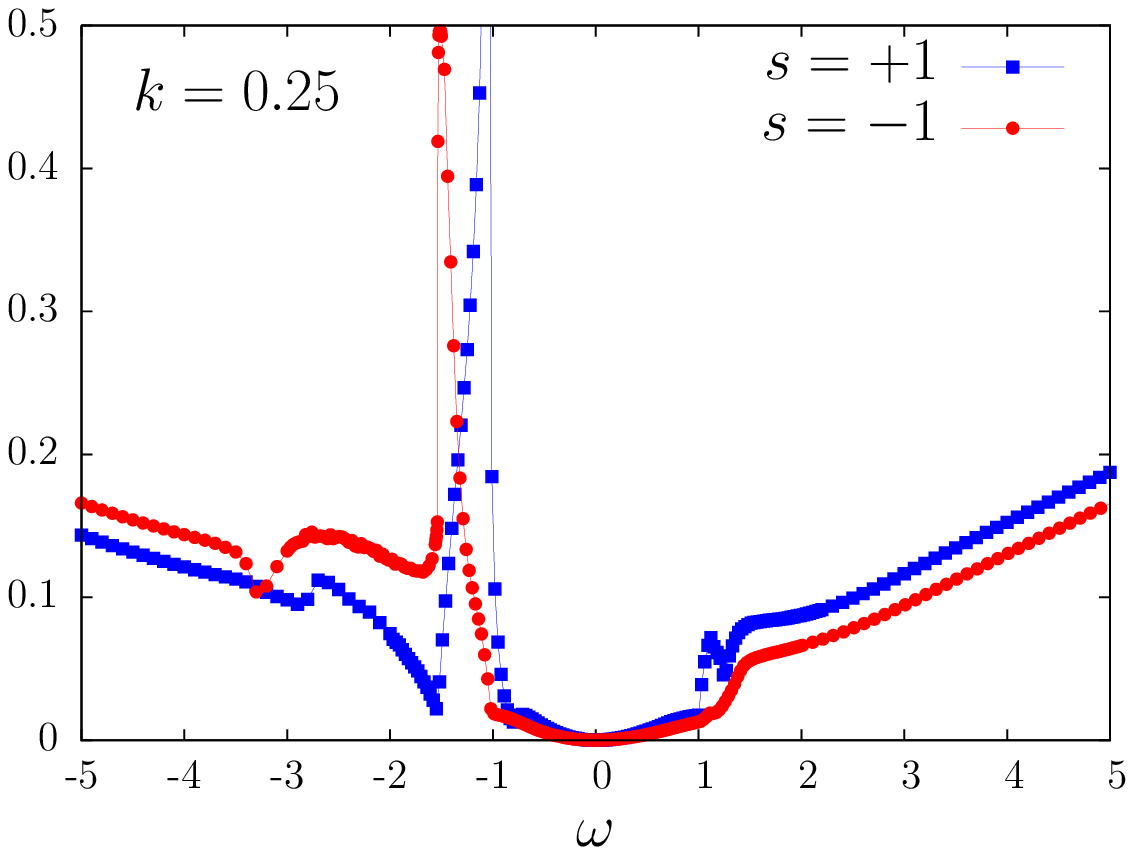}\\
\includegraphics[width=0.50\linewidth]{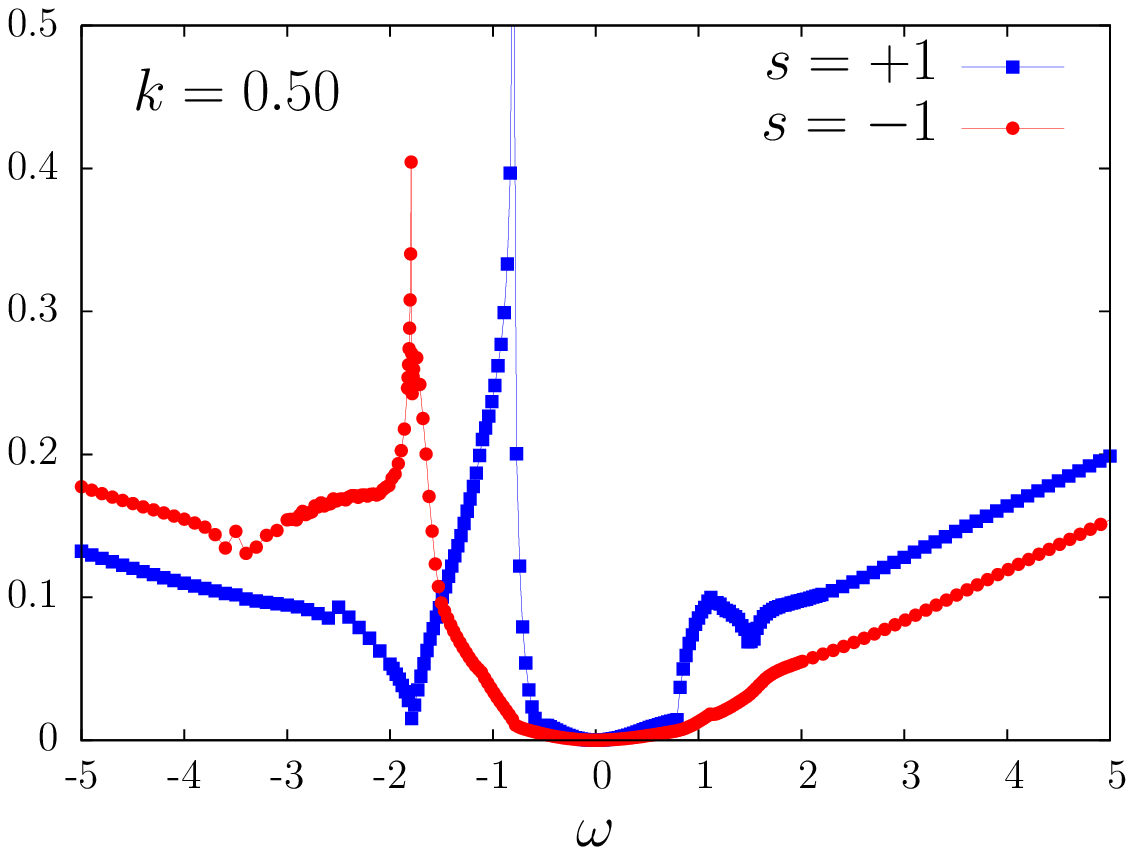}&
\includegraphics[width=0.50\linewidth]{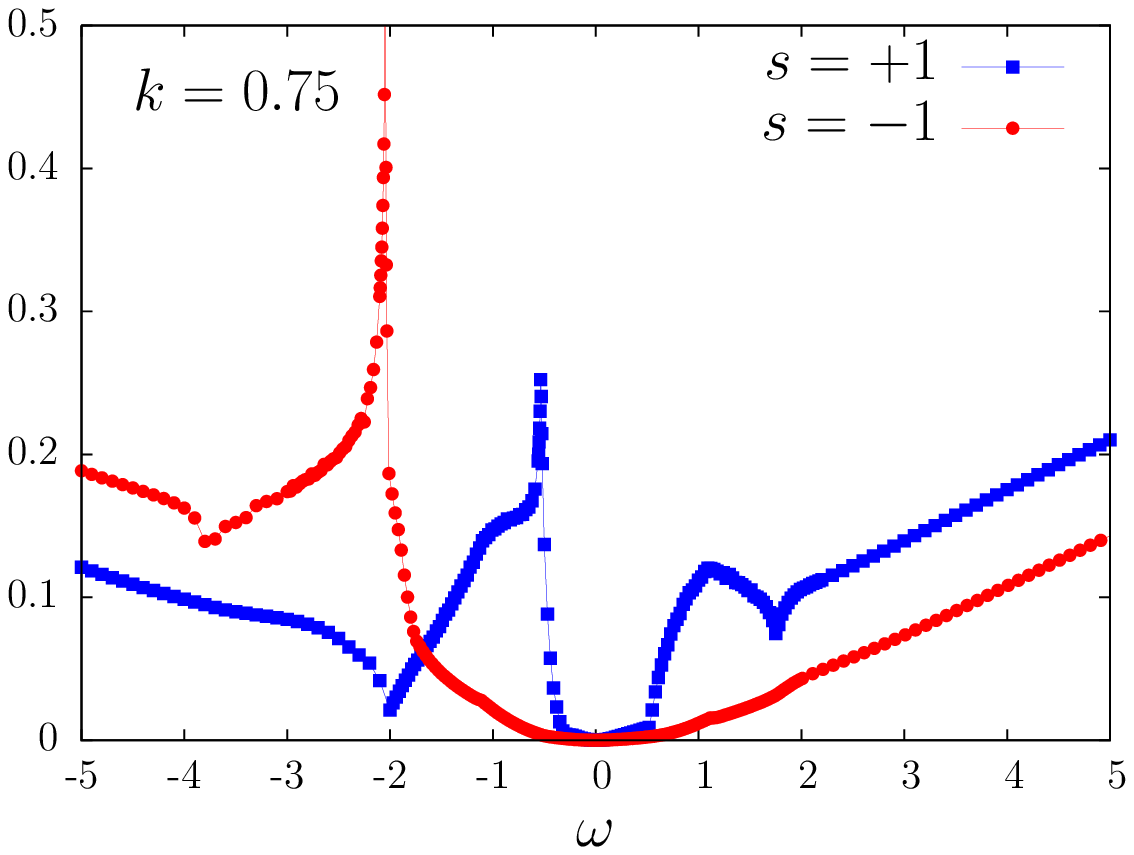}
\end{tabular}
\caption{(Color online) The absolute value $|\Im m [\Sigma_s({\bm k}, \omega)]|$ 
of the imaginary part of the RPA quasiparticle self-energy (in units of $\varepsilon_{\rm F}$)
of an n-doped system as a function of energy $\omega$ for $k=0,0.25,0.50$, and $0.75$ and for $\alpha_{\rm gr}=2$.\label{fig:three}}
\end{center}
\end{figure}

\noindent {\em Dirac Quasiparticle Decay}---
In Fig.~\ref{fig:three} we plot our results for the imaginary part of the Dirac sea self-energy for a
series of ${\bm k}$ values (these results are for a n-doped $\alpha_{\rm gr}=2$ system). 
In the RPA  
\begin{widetext} 
\begin{equation} 
\label{eq:res_computation}
\Im m [\Sigma_{s}({\bm k},\omega)]= \sum_{s'} 
\int \frac{d^2 {\bm q}}{(2\pi)^2}~v_q~\Im m[\varepsilon^{-1}({\bm q},\omega-\xi_{s'}({\bm k}+{\bm q})) 
\left[\frac{1+ ss'\cos{(\theta_{{\bm k},{\bm k}+{\bm q}})}}{2}\right]
\left[\Theta(\omega-\xi_{s'}({\bm k}+{\bm q}))-\Theta(-\xi_{s'}({\bm k}+{\bm q})) \right] 
\end{equation}
\end{widetext}
where $s,s'=\pm 1$ are band indices, $v_q=2\pi e^2/(\epsilon q)$ is the 2D Coulomb interaction,
$\varepsilon({\bm q},\omega)=1-v_q \chi^{(0)}({\bm q}, \omega)$ is the RPA dielectric function, 
and $\Theta(x)$ is the Heaviside step function.
The two factors in square brackets on the right-hand-side of Eq.~(\ref{eq:res_computation}) express respectively the 
influence of chirality and Fermi statistics on the decay process.  Note that $\Sigma_s$ depends on the band-index $s$ only
through the chirality factor.  For $\omega > 0$ and fixed ${\bf q}$, the RPA decay process represents scattering of an 
electron from momentum ${\bm k}$ and energy $\omega$ to ${\bm k}+{\bm q}$ and $\xi_{s'}({\bm k}+{\bm q})$, with all 
energies in Eq.~(\ref{eq:res_computation}) measured from the Fermi energy.  Since the Pauli exclusion principle requires that the final state
is unoccupied, it must lie in the conduction band, {\it i.e.} $s'=+1$.  Furthermore since the Fermi sea is initially in its 
ground state, the quasiparticle must lower its energy, {\em i.e.} $\xi_{s'} < \omega$ -- electrons decay by going down in energy. 
For $\omega < 0$, the self-energy expresses the decay of holes inside the Fermi sea, which scatter to a final state, by exciting
the Fermi sea.  In this case the final state must be occupied so both band indices are allowed for $s'$, and 
energy conservation requires that holes decay by moving up in energy.  Since photoemission measures the properties of 
holes produced in the Fermi sea by photo ejection, only $\omega < 0$ is relevant for this experimental probe.  
Note however that because of the particle-hole symmetry properties of the Dirac model 
$\Sigma_{s}({\bm k},\omega)$ in an n-doped system is identical to $\Sigma_{-s}({\bm k},-\omega)$ in 
a system with the opposite doping.  Our results for $\Im m [\Sigma_{s}({\bm k},\omega)]$ for $\omega >0$,
therefore specify the ARPES spectra of p-doped graphene. 
 
We remark that because interaction and band energies in graphene's Dirac model both scale inversely with 
length, $\Im m [\Sigma_{s}({\bm k},\omega)] = v k_{\rm F} F(\omega/v k_{\rm F}, k/k_{\rm F})$.  For large $|x|$, 
$F(x,y) \to -\pi  \alpha^2_{\rm gr} \ell(\alpha_{\rm gr})|x|/(64 g)$, where 
\begin{equation}
\ell(\alpha_{\rm gr})=\int_0^{1}dx~\frac{x\sqrt{1-x}}{1-x+\pi^2\alpha^2_{\rm gr} x^2/256}
\end{equation}
[$\ell(0)=4/3$, $\ell(2)\simeq 0.655124$]. This implies that 
for $|\omega| \gg v k_{\rm F}$, the decay rate in a doped system ($\Im m [\Sigma_{s}({\bm k},\omega)]$) 
approaches that of an undoped system (the Fermi energy $\varepsilon_{\rm F} = v k_{\rm F}$ is used as the energy
unit and $k_{\rm F}$ as the unit of wavevector in all plots and in the remaining sections of this Letter). 
As we will see, however, the doped system properties are quite different from those of an undoped system up to energy scales several
times larger than the Fermi energy, particularly so near the Dirac ($k=0$) point. 

In explaining the spectra plotted in Fig.~\ref{fig:three} we start with the $k=0$ case 
($\Im m[\Sigma_+(0,\omega)]=\Im m[\Sigma_-(0,\omega)]$) for which the final state energy 
is $\xi_{s'}({\bm q}) =s'q -1$.  For $\omega > 0$
the final state must be unoccupied 
so that $s'=+1$; $q$ is restricted to those values larger than $1$ for which the 
the Dirac sea excitation energy $\Omega(q)= \omega +1 - q$ is positive.    
Comparing with Fig.~\ref{fig:two} we see that $\Im m[\Sigma_+(0,\omega)]$ vanishes 
like $\omega^2$ for $\omega \to 0$, a universal property of normal Fermi liquids.
In the RPA 
\begin{equation}
\Im m[\Sigma_+(0,\omega \to 0)]=-\frac{\sqrt{3}}{8g}\frac{\alpha^2_{\rm gr}}{(1+\alpha_{\rm gr})^2}\omega^2\,.
\end{equation}
The sharp increase in $\Im m[\Sigma_+(0,\omega)]$ which occurs 
at $\omega \sim 1.2$ reflects the onset of plasmon emission.  Note that these plasmons 
at $q >1$ remain well defined excitations well into the interband particle-hole continuum.
For $\omega <0$, relevant to ARPES in n-doped graphene, both conduction and valence band 
final states occur and transitions are allowed if the transition energy $\Omega(q)= |\omega|-1 + s'q$ is positive and 
the final hole state is occupied, {\it i.e.} $q < 1$ for $s'=+$.  Note that for final states in the 
conduction band, the transition energy is close to the plasmon excitation energy over a wide range of 
$q$ values.  This property leads to much stronger plasmon features for $\omega < 0$ than for $\omega > 0$. 
$\Im m [\Sigma_+(0,\omega)]$ diverges for $\omega \sim -1.3$, the value at which $\Omega(q)$ for $s'=+1$ 
is tangent to the plasmon dispersion line illustrated in Fig.~\ref{fig:two}.  In this case, the hole scatters into 
a resonance consisting of a quasiparticle strongly coupled to plasmon excitations, a plasmaron~\cite{Hedin,Jalabert}.
At finite $k$, the conduction and valence band $\Im m [\Sigma_{s}({\bm k},\omega)]$ plasmaron peaks broaden and separate,
because of the dependence on scattering angle of $\xi_{s'}({\bm k}+{\bm q})$ and because of chirality factors which 
emphasize ${\bm k}$ and ${\bm q}$ in nearly parallel directions for conduction band states and 
${\bm k}$ and ${\bm q}$ in nearly opposite directions for valence band states.  Because of the chirality factor
the conduction band plasmaron moves up in energy approximately as $vk$ while the valence band plasmaron moves down. 

\begin{figure}[t]
\begin{center}
\tabcolsep=0cm
\begin{tabular}{cc}
\includegraphics[width=0.50\linewidth]{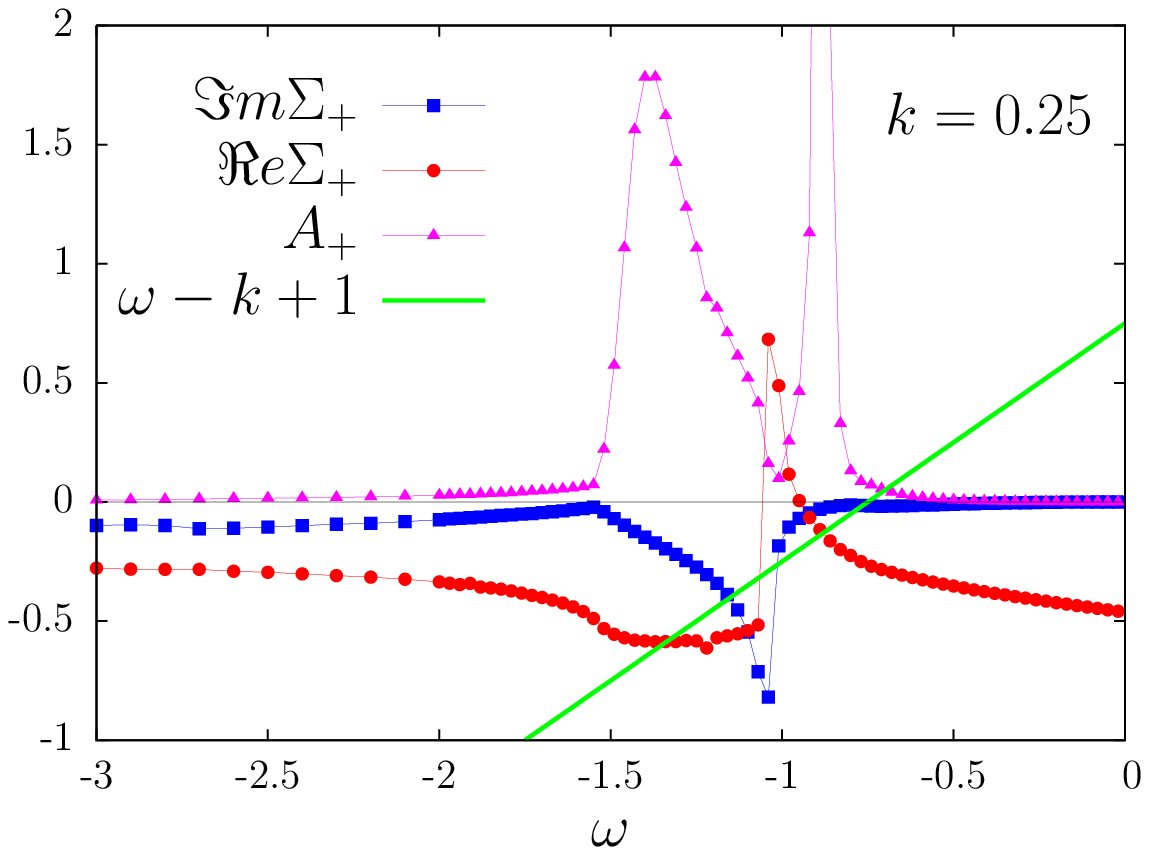}&
\includegraphics[width=0.50\linewidth]{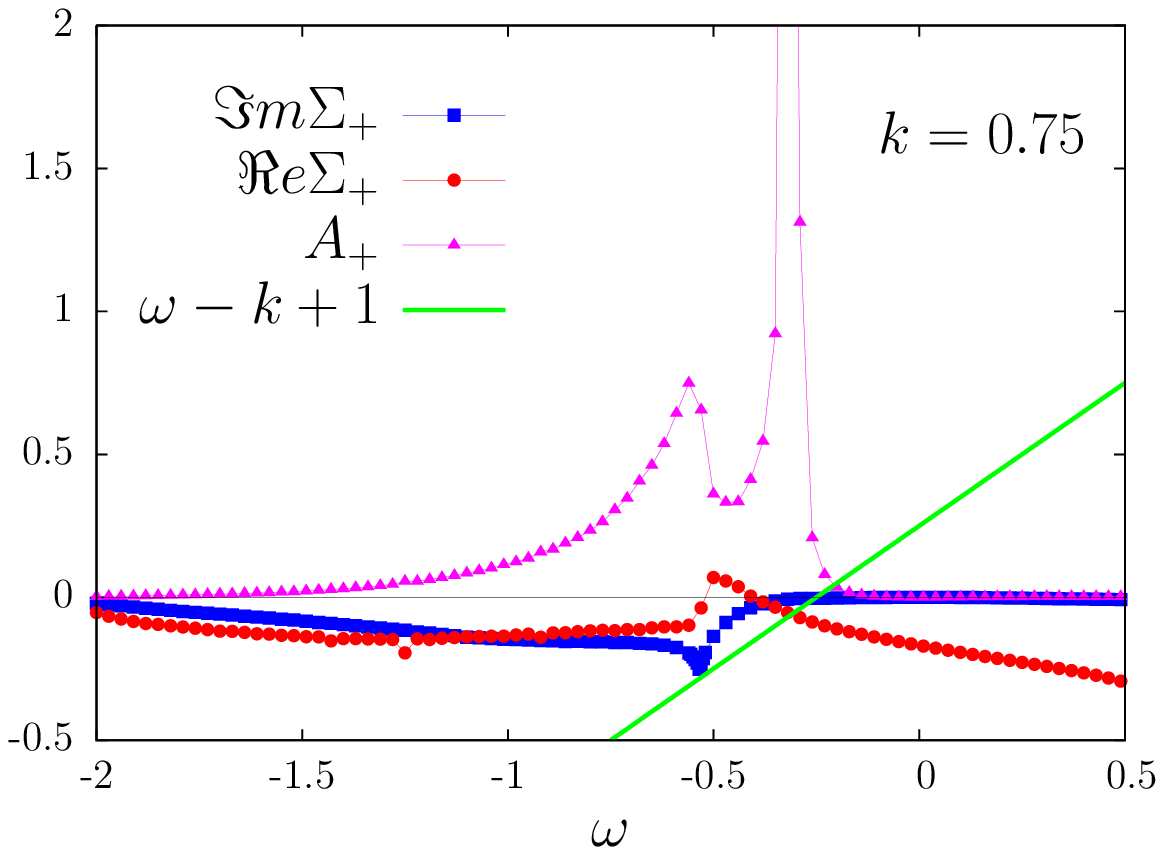}\\
\end{tabular}
\caption{(Color online) $\Re e [\Sigma_+({\bm k}, \omega)]$, $\Im m [\Sigma_+({\bm k}, \omega)]$, and spectral function
${\cal A}_+({\bm k}, \omega)$ for $k=0.25$ and $k=0.75$.  The band energy and $\Re e \Sigma_+$ are measured from the 
band $\varepsilon_{\rm F}$ and interaction [$\Sigma_{+}(k_{\rm F},\omega=0)$] contributions to the chemical potential.
\label{fig:four}}
\end{center}
\end{figure}

\noindent
{\em ARPES Spectra}---We are now prepared to discuss the ARPES spectra in Fig.~\ref{fig:one}.  
ARPES measures the 
wavevector dependent electron spectral function
\begin{equation}\label{eq:spectral}
{\cal A}_s({\bm k},\omega)=\frac{1}{\pi}\frac{|\Im m \Sigma_s({\bm k},\omega)|}{[\omega-\xi_{s}({\bm k})-\Re e \Sigma_s({\bm k},\omega)]^2+[\Im m \Sigma_s({\bm k},\omega)]^2}\,.
\end{equation}
Near the Fermi energy the spectral function consists of a narrow Lorentzian centered at the energy $E$ which solves the Dyson
equation for the quasiparticle energy:
\begin{equation} 
E = \xi_{s}({\bm k})+\Re e \Sigma_s({\bm k}, E).
\end{equation} 
For small $k$ we see in Fig.~\ref{fig:four} that there are two solutions to this equation, which are shifted from the 
bare quasiparticle energy and the plasmaron energy because of the strong energy-dependence of $\Re e \Sigma_s$ near the 
$\Im m \Sigma_s$'s plasmaron peak.  These shifts can be understood as following from level repulsion between a 
bare particle resonance and a plasmaron resonance.  We also note in Fig.~\ref{fig:four} that $\Re e \Sigma_s$ has a 
negative contribution which is present at the Fermi energy and persists over a wide regime of energy.  This contribution 
is due to exchange and correlation interactions of quasiparticles near the Fermi energy with the negative energy sea.
As explained~\cite{ourprl,poliniSSC,footnote} previously, this effect produces a nearly rigid shift in the band 
energies which is increasingly negative further below the Fermi energy, increasing the band 
dispersion and the quasiparticle velocity.  In Fig.~\ref{fig:four} we show that for larger $k$, the plasmaron feature in
$\Im m \Sigma_s$ is broadened sufficiently to remove the plasmaron solution of the quasiparticle Dyson equation.  
In this limit, the spectral function has only a weak plasmon satellite on top of the main quasiparticle peak.

The full ARPES spectral in Fig.~\ref{fig:one} can be understood by applying considerations like
those explained above to both bands and over the full range of $k$.  Because the negative exchange-correlation shift in the 
quasiparticle energy gets larger as $k \to 0$ in the conduction band, the energy splitting between the sharpening plasmaron
resonance and the band-like quasiparticle decreases and their coupling increases.  In addition, because the slope difference 
between $E-\xi_s({\bm k})$ and $\Re e \Sigma_s({\bm k}, E)$ is smaller at the plasmaron Dyson equation solution than at the 
quasiparticle Dyson equation solution, the plasmaron assumes the larger fraction of the 
spectral weight. The detailed behavior near $k =0 $ is sensitive to the value of $\alpha_{\rm gr}$.  
For $\alpha_{\rm gr}=2$, the avoided crossing between quasiparticle and plasmaron solutions occurs 
near $k = 0$.  Because the higher energy valence band peak has little spectral weight when it crosses 
the Fermi energy, the overall spectrum can appear to have an energy gap.

Because electron-electron interaction effects occur on the $v k_{\rm F}$ energy scale they can be separated from electron-phonon
interaction effects experimentally by varying carrier density.  
Although the RPA is not exact, it provides a good starting point for interpreting the 
influence of electron-electron interactions on the ARPES spectra of graphene sheets.
As we have shown, the downward shift of valence band states compared to conduction band 
states and plasmaron-quasiparticle level repulsion and weight transfer effects 
must be accounted for to interpret ARPES data.  Features in the data which might otherwise 
be interpreted as extrinsic effects related to disorder or interactions with substrates 
can have interesting and subtle intrinsic origins.  Close comparison between theory and experiment 
and further materials progress will augment the power of ARPES experiments.   

\noindent
{\em Acknowledgment ---}
This work has been supported by the Welch Foundation, by the Natural Sciences and Engineering Research Council of Canada,
by the Department of Energy under grant DE-FG03-02ER45958, and by the National Science Foundation under grant DMR-0606489.
We thank Alessandra Lanzara, Steve Louie, and Andrea Tomadin for helpful discussions.


\begin{thebibliography}{99}
\bibitem{results}  This information is available in numerical form upon request to the authors.
\bibitem{footnote} 
$\Re e \Sigma_s$ is weakly dependent on the massless Dirac model's 
ultraviolet cutoff $\Lambda$~\cite{poliniSSC,ourprl}.  For this reason the spectral function ${\cal A}_{s}({\bm k},\omega)$ is not 
exactly a function of only $k/k_{\rm F}$ and $\omega/\varepsilon_{\rm F}$.  This caveat is 
of little practical signifigance however since deviations are very small over the 
two to three orders of magnitude of graphene sheet charge density which 
lies the window between the low density cutoff $\sim  10^{11} {\rm cm}^{-2}$ 
below which disorder plays a strong role and the high density 
cutoff $\sim 10^{14} {\rm cm}^{-2}$ imposed by fundamental gating and doping limitations. The numerical results shown in this work have been calculated with $\Lambda=10^2$.
\end{thebibliography}
\end{document}